\documentclass{article}

\usepackage{PRIMEarxiv}

\usepackage[utf8]{inputenc} % allow utf-8 input
\usepackage[T1]{fontenc}    % use 8-bit T1 fonts
\usepackage{hyperref}       % hyperlinks
\usepackage{url}            % simple URL typesetting
\usepackage{booktabs}       % professional-quality tables
\usepackage{amsfonts}       % blackboard math symbols
\usepackage{nicefrac}       % compact symbols for 1/2, etc.
\usepackage{microtype}      % microtypography
\usepackage{lipsum}
\usepackage{fancyhdr}       % header
\usepackage{graphicx}       % graphics

\newcommand{\degree}{^{\circ}}
\usepackage{amssymb,amsthm,amsmath}
\usepackage[T1]{fontenc}
\usepackage{mathptmx}
\usepackage{etoolbox}
\usepackage{array}
\usepackage{multirow}
\usepackage{xcolor}
\graphicspath{{media/}}     % organize your images and other figures under media/ folder

\title{Resolving the band alignment of InAs/InAsSb mid-wave-infrared type-II superlattices}

\author{
  Michał Rygała \\
  Wrocław University of Science and Technology\\
  \texttt{michal.rygala@pwr.edu.pl}\\
  \And
  Julian Zanon \\
  Eindhoven University of Technology\\
  \And
  Andreas Bader \\
  Julius-Maximilians-Universität Würzburg\\
  \And
  Tristan Smołka \\
  Wrocław University of Science and Technology\\
  \And
  Fabian Hartmann \\
  Julius-Maximilians-Universität Würzburg\\
  \And
  Sven H{\"o}fling \\
  Julius-Maximilians-Universität Würzburg\\
  \And
  Michael E. Flatté \\
  University of Iowa\\
  Eindhoven University of Technology\\
  \And
  Marcin Motyka \\
  Wrocław University of Science and Technology\\
  }  

\begin{document}
\maketitle

\begin{abstract}
In this work, three InAs/InAs$_{0.65}$Sb$_{0.35}$ superlattices with different periods were investigated using photoluminescence and photoreflectance measurements and their band structure was simulated using a 14 bulk-band  \textbf{k}$\cdot$\textbf{p} model. The structures were studied by analyzing the evolution of the spectral features in temperature and excitation power to determine the origin of optical transitions. After identifying which of these are related to the superlattice mini-bands, a rich collection of observed higher-order optical transitions was compared with  refractive-index calculations. This procedure was used to adjust the parameters of the theoretical model, namely the bowing parameters of the InAsSb valence band offset and bandgap. It was also shown that the spectroscopy of the higher-order states combined with numerical modeling of the refractive index is a powerful tool for improvement of the material parameters, presenting a new approach to material studies of advanced semiconductor heterostructures.
\end{abstract}

\section{INTRODUCTION}
In the last 60 years of searching for an efficient infrared detector, many solutions were reported, with their own advantages and disadvantages. %here some more reviews should be cited, not just from co-authors on paper but from other groups, for politeness
The recent development of highly efficient growth procedures for \(A^{III}B^{V}\) semiconductor alloys \cite{Rogalski2012, Rogalski2006} is closing the technological gap between detectors based on low-dimensional structures and widely used HgCdTe alloys. \cite{Smith1987,Rehm2005,Maimon2006,Plis2006,Nguyen2008,Kopytko2024} Despite the overall lower quantum efficiency of \(A^{III}B^{V}\) semiconductor-based detectors in comparison to \(A^{II}B^{VI}\), they are more mechanically stable due to the covalent bonding of the atoms \cite{Rogalski2017} and are generally less hazardous to health and the environment. \cite{Madni2017,Kopytko2022,Nordin2022}

Primary interest has focused on two material systems with type-II superlattice (T2SL) architecture - InAs/GaSb and InAs/InAsSb. Both of these materials were widely investigated, with overall optical properties favoring  the InAs/GaSb SLs and electronic properties favoring the InAs/InAsSb SLs. \cite{Rogalski2020,Alshahrani2022} The absorption coefficient of the "Ga-free" SLs is lower, mainly due to the unfavorable overlap of the electron and hole states within the structure. \cite{Mailhiot1989,Smith1990}
However, they exhibit much longer Shockley-Read-Hall (SRH) carrier lifetimes than their Ga-containing counterparts \cite{Belenky2011, Steenbergen2011, Olson2012, Lin2015, Aytac2016, DeWames2019}, which has been attributed to the removal of gallium from the structure . \cite{Jóźwikowska2019} The detectivity of the InAs/InAsSb-based detectors influenced by both of these effects has been reported to be approaching the ultimate efficiency reported for the HgCdTe and defined by semi-empirical curves, such as "Rule-07", "Law 19", or "Rule-22". \cite{Kopytko2024}

In InAs/InAsSb alloys the energy difference between the bottom of the valence and conduction bands can be modified by the composition of the InAsSb layer and the thickness of the single superlattice period. The lattice constant of the alloy follows Vegard's law, therefore it varies linearly with the composition, whereas the approximation of the bandgap uses an additional bowing parameter. \cite{Svensson2012} Moreover, to fully describe the transition energies between the confined states within the InAs/InAsSb T2SL it is crucial to know the energy difference between the bottoms of the valence bands of InAsSb and InAs. This is usually described by the valence band offset parameter (VBO), which has been presented to be composition-dependent \cite{Vurgaftman2001} but also temperature-dependent \cite{Steenbergen2012,Manyk2018}, leading to a plethora of speculations in terms of the InAsSb valence band position and its bandgap. It has also been shown that the type of band discontinuity can vary between type I, IIa, IIb, and III (broken gap alignment) \cite{Steenbergen2011}.

In this work, we present the results of the experimental and numerical investigation of InAs/InAs\(_{0.65}\)Sb\(_{0.35}\) T2SLs on a GaSb substrate with different period lengths. The analysis of the optical response of the structures was performed using a 14 bulk band \textbf{k}$\cdot$\textbf{p}  model to perform calculations of confined states. The InAsSb VBO and bandgap bowing parameters were modified during model optimization in order to match the results of refractive index calculations in the presence of external carriers and photomodulated reflectance (PR) spectra.

\section{MATERIALS AND METHODS}
\subsection{Sample growth}
Investigated samples consisted of InAs/InAs\(_{0.65}\)Sb\(_{0.35}\) superlattices with varying periods \(P\) of \(5\), \(6\) and \(8\ \textnormal{nm}\) . The structures were grown using molecular beam epitaxy on n-type doped GaSb:Te substrates with doping concentration of \(10^{18}~\textnormal{cm}^{-3}\). Growth was initiated with a deoxidization step at \(560~\degree \textnormal{C}\) which was followed by the growth of an undoped 200 nm GaSb buffer layer at a substrate temperature of \(500~\degree \textnormal{C}\). Subsequently, the growth temperature was reduced to \(430~\degree \textnormal{C}\) for the growth of the T2SL. The thicknesses of the individual SL layers were chosen to achieve overall strain compensation on the GaSb substrate, therefore making the system lattice-matched. For the \(P=5~\textnormal{nm}\) sample, the InAs and InAsSb thicknesses were 3.71 and 1.23 nm , respectively. For \(P=6~\textnormal{nm}\), the layer thicknesses were 4.45 (InAs) and 1.55 nm (InAsSb), and for \(P=8~\textnormal{nm}\): 5.94 (InAs) and 2.06 nm (InAsSb). \(100\) \( (P=8, 6~\textnormal{nm})\) or \(150\) \((P=5~\textnormal{nm})\) periods of the SLs were grown and the structures capped with a 50 nm undoped GaSb layer.\par
\subsection{Fourier spectroscopy}
Photoluminescence (PL) and photoreflectance (PR) measurements were performed on a Vertex 80v Fourier spectrometer using a 660 nm excitation beam. The samples were kept in a liquid helium cryostat at a stable 10 K temperature. The emitted or reflected light was collected by an MCT detector in a step scan measurement mode correlated with a chopper by a homodyne for a 300 ms integration time per interferometer mirror step. This experimental setup has proven useful for investigating weak transitions in semiconductor nanostructures designed for the infrared spectral region. \cite{Rygala2025,Rygala2021,Motyka2010,Motyka2011}\par
\subsection{Numerical model}
In order to obtain the VBO between the InAs and InAsSb layers, the photoreflectance spectra for the three samples with different periods were compared with the change in the refractive index (i.e., $\Delta$n) calculated using the 14 bulk band \textbf{k}$\cdot$\textbf{p} model from QuantCAD's software called CADtronics \cite{lau2004electronic}. 

The 14 bulk band \textbf{k}$\cdot$\textbf{p} model includes in its basis two antibonding $s$ states, six valence bonding $p$ states, and six antibonding $p$ states. The inclusion of these states takes into account the inversion asymmetry present in zincblende materials (such as InAs or InSb), and provides a more accurate treatment for the heavy-hole band mass \cite{krier2007mid} as the momentum \textbf{k} is further away from \textbf{k}$=\boldsymbol{0}$. Furthermore, the model used here also considers energy shifts due to lattice-mismatch strain between the two different semiconductors. 

The VBO is obtained after an optimization procedure, where the transitions obtained in $\Delta$n are compared to the ones from the photoreflectance. As detailed in {\color{blue}{Supplementary Materials}}, the refractive index n$(\text{E};\rho)$ for a given carrier density $\rho$ is calculated as a function of energy $\text{E}$ using the absorption $\alpha(\text{E};\rho)$, calculated using the complex dielectric function $\varepsilon(E)$. Following that, the change in the refractive index $\Delta$n is obtained with $\Delta$n = n$(\text{E};0)$ - n$(\text{E};\rho)$. While comparing the transitions energies between the photoreflectance and $\Delta$n, a careful analysis is carried out considering different values for the valence band bowing b$_{\text{v}}$ and the gamma gap bowing b$_{\text{g}}$ from the InAsSb layer. b$_{\text{v}}$ and b$_\text{g}$ are related to the valence band E$_{\text{v}}$ and gap 
E$_{\text{g}}$ energies of the InAsSb layer at the gamma point, as presented in Eq.(\ref{eq:InAsSb_energies}). 
\begin{equation}
\begin{aligned}
    \text{E}_{\text{v(g)}}(\text{InAs}_{1-x}\text{Sb}_{x}) &= (1-x)\text{E}_{\text{v(g)}}(\text{InAs})+ x\,\text{E}_{\text{v(g)}}\,(\text{InSb})\\
    &-\text{b}_{\text{v(g)}}\,x(1-x)\,\ \label{eq:InAsSb_energies}
\end{aligned}
\end{equation}

\section{RESULTS}
\begin{figure}
    \centering    
    \includegraphics[width=1.0\linewidth]{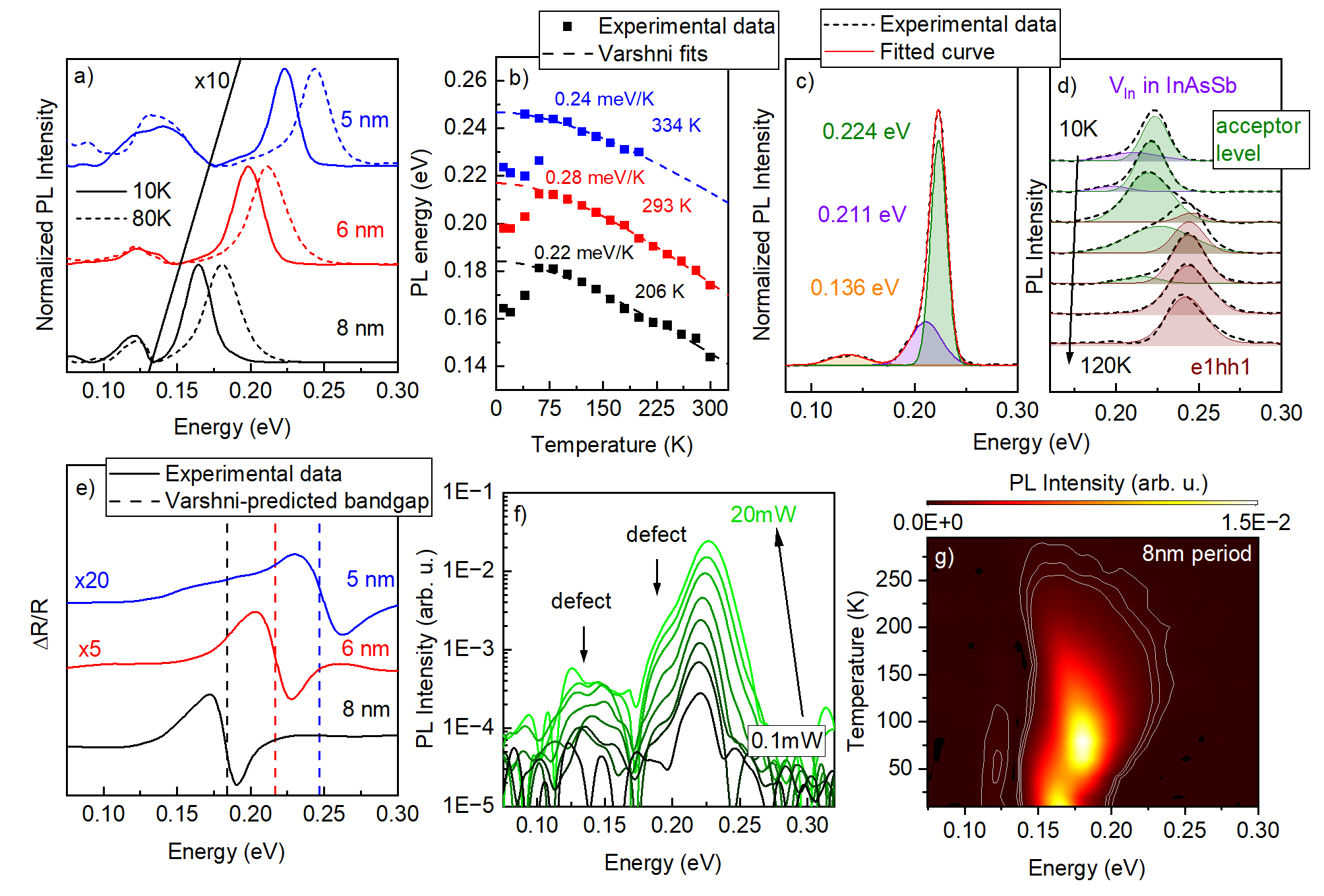}
    \caption{Optical properties of the InAs/InAs\(_{0.65}\)Sb\(_{0.35}\) superlattices. a) Normalized photoluminescencence spectra of the samples at 10 K (solid line) and 80 K (dashed line) temperature shown in blue, red and black for sample with 5 nm, 6 nm and 8 nm period, respectively. b) Energies of the dominating photoluminescence emission fitted with Varshni equation shown in blue, red and black for samples with 5 nm, 6 nm and 8 nm period, respectively, with the parameters of the fitted function - \(\alpha\) and \(\beta\). c) Close up of the emission lineshape of 5 nm period sample at 10 K temperature with the red fitted curve comprising of three distinct gaussian peaks with the green, violet and orange colours representing the optical transitions originating from acceptor level, a defect state corresponding to indium vacancy and an unknown confined state. d) Temperature evolution of the photoluminescence signal from the sample with 5 nm period with fitted gaussian peaks colour coded as in the panel c and additional wine colour representing a fundamental transition from the first confined states within the conduction mini-band to the valence mini-band. e) Photoreflectance spectra of the fundamental transitions shown in blue, red and black for the samples with 5 nm, 6 nm and 8 nm period, respectively, with the bandgap energies predicted by fitting the Varshni equation to higher-temperature energies of the dominating photoluminescence emission, presented as dashed lines with colours corresponding to the samples. f) Power-dependent photoluminescence spectra of the sample with 5 nm period at 10 K in logarithmic scale, denoting the defect states within the spectra. g) Temperature evolution of photoluminescence signal from 8 nm period sample showing the maximum of signal intensity at around 80 K.}
    \label{fig:experimental figure 1}
\end{figure}

The optical properties of the samples are presented in Figure \ref{fig:experimental figure 1}. All samples exhibit a shift of the emission to higher energies caused by changing the temperature from 10 to 80 K, as shown in Figure \ref{fig:experimental figure 1}a. The energy of the main photoluminescence signal also slightly deteriorates at temperatures below 80 K from the bandgap, predicted by fitting the energies of high-temperature range emission with the Varshni equation, as shown in Figure \ref{fig:experimental figure 1}b. The non-monotonic behavior of the emission energy in varying temperature referred to as "S-shape" has been previously investigated and attributed to carrier localization due to mid-gap trap states.\cite{Steenbergen2016, Murawski2023} The fitting procedure of high-temperature side yielded the values of parameters \(\alpha\) and \(\beta\) in the range of \(0.22 - 0.28\,~\)meV/K and \(206 - 334\,\)~K, respectively. While the \(\alpha\) parameter is in agreement with the values reported in literature \cite{Steenbergen2016, Manyk2019, Smolka2022, Murawski2023}, the \(\beta\) is slightly higher, possibly due to the lack of emission data from the fundamental transition in low temperature regime. The bandgap energies predicted by the fitting were equal to \(0.246\), \(0.217\) and \(0.185\,~\)eV for 5, 6 and 8 nm period sample, respectively. 

To further explore this unusual behavior, we performed a detailed analysis of emission lineshape which was shown for 5 nm period sample in Figure \ref{fig:experimental figure 1}c.
At 10 K, it is possible to distinguish three separate signals comprising the photoluminescence envelope at energies 0.224, 0.211 and 0.136 eV. After the temperature of the sample was increased to 40 K, another signal (presented in wine color in the Figure \ref{fig:experimental figure 1}d) emerged at a higher energy corresponding to the e1hh1 energy predicted by the Varshni equation. It later dominates the photoluminescence spectra at the cost of the signal denoted in green colour. By comparing these findings with literature \cite{Murawski2024,Krishnamurthy2017} we identified the green signal as an acceptor level-related transition, while the violet is predicted to arise from a mid-gap trap level related to an indium vacancy in InAsSb layer of the superlattice. This type of structural defect is predicted to be one of the main limitations in terms of efficiency of InAsSb-based infrared detectors. \cite{Aytac2016} \par The defect-related origin of the dominant signal at 10 K is also supported by photoreflectance measurements as presented in Figure \ref{fig:experimental figure 1}e. The resonance energies of the photoreflectance spectra match the predicted low-temperature bandgap energies obtained from the Varshni equation. The main low-temperature emission must therefore arise from a bound level characterized by low density of states, which would not be observable through absorption-based experiments. These findings were also supported by power-dependent photoluminescence measurements shown in Figure \ref{fig:experimental figure 1}f. The power dependence of the main low-temperature emission peak is sub-linear which indicates that the emission is caused by carrier recombination to a confined (defect) state or related to donor-acceptor levels. Increasing the temperature to 77 K resulted in a linear relation between excitation power and photoluminescence intensity, expected for high-intensity regime excitation of type-II quantum system. \cite{Chiu2002,Alshahrani2023,Rogowicz2022} Defect-related emission at lower energies was also investigated and it resulted in an exponent value close to \(1/2\) that is characteristic for a bound-to-defect optical transitions. \cite{Deng2019} The value remained constant for 77 K, which indicates a high activation energy of this defect. Further study of the relation between the emission and the temperature of the sample as presented in Figure \ref{fig:experimental figure 1}g provides more information about the origin of the emission. The intensity is highest at a temperature in which the energy of the transition is in agreement with the Varshni-predicted bandgap. The defect-related emission is observable up to 150 K.\par
\begin{figure}
    \centering
    \includegraphics[width=0.5\linewidth]{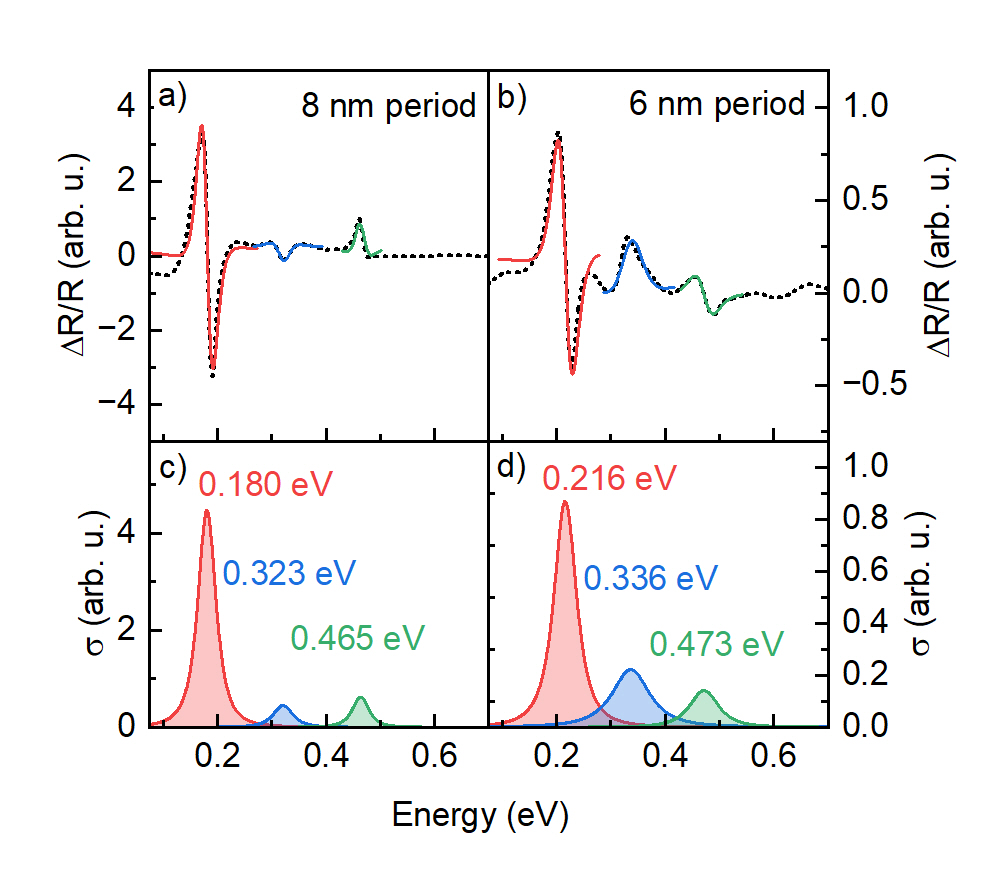}
    \caption{Results of the photoreflectance experiment. a,b) Photoreflectance spectra of the sample with 8 nm (panel a)), and 6 nm (panel b) period at 10 K temperature with fitted curves using Equation \ref{PR resonance}. c,d) Moduli calculated with the Equation \ref{PR modulus} using the parameters from the fitting procedure of the photoreflectance response for sample with 8 nm (panel c)) and 6 nm period (panel d)).}
    \label{fig:experimental figure 2}
\end{figure}

The photoreflectance response of the 8 and 6 nm samples for a broader range of energy are presented in Figure \ref{fig:experimental figure 2} (panel a and b, respectively). Apart from the fundamental optical transitions at 0.180 eV (8nm sample) and 0.216 eV (6 nm sample), there are other features corresponding to higher-order states within the superlattice. The sample with a 5 nm period also exhibited higher-order features and is presented in the later parts of the manuscript. The resonances of the reflectivity spectra were fitted with the function described by the equation: 
\begin{equation}
    \frac{\Delta R}{R} = Re{\sum_{j=1}^{n}[C_j^{i\theta_j}(E-E_j+i\Gamma_j)]^{-m_j}}
    \label{PR resonance}
\end{equation}
where \(n\) is the number of optical transitions, \(C_j\) and \(\theta_j\) are amplitude and phase parameters, \(E_j\) and \(\Gamma_j\) are energy and broadening of the transition and \(m_j\) describes the lineshape of the modulation accounting for the type of transition (excitonic, free-carrier, etc.)\cite{Shanabrook1987}. For clarity of presentation, the transitions were also shown as moduli in Figure \ref{fig:experimental figure 2} (panels c and d) described by the equation:
\begin{equation}
    \sigma = \frac{|C|}{[(E-E_0)^{2}+\Gamma^{2}]^{\frac{m}{2}}}
    \label{PR modulus}
\end{equation}

\begin{figure}
    \centering
    \includegraphics[width=1.0\linewidth]{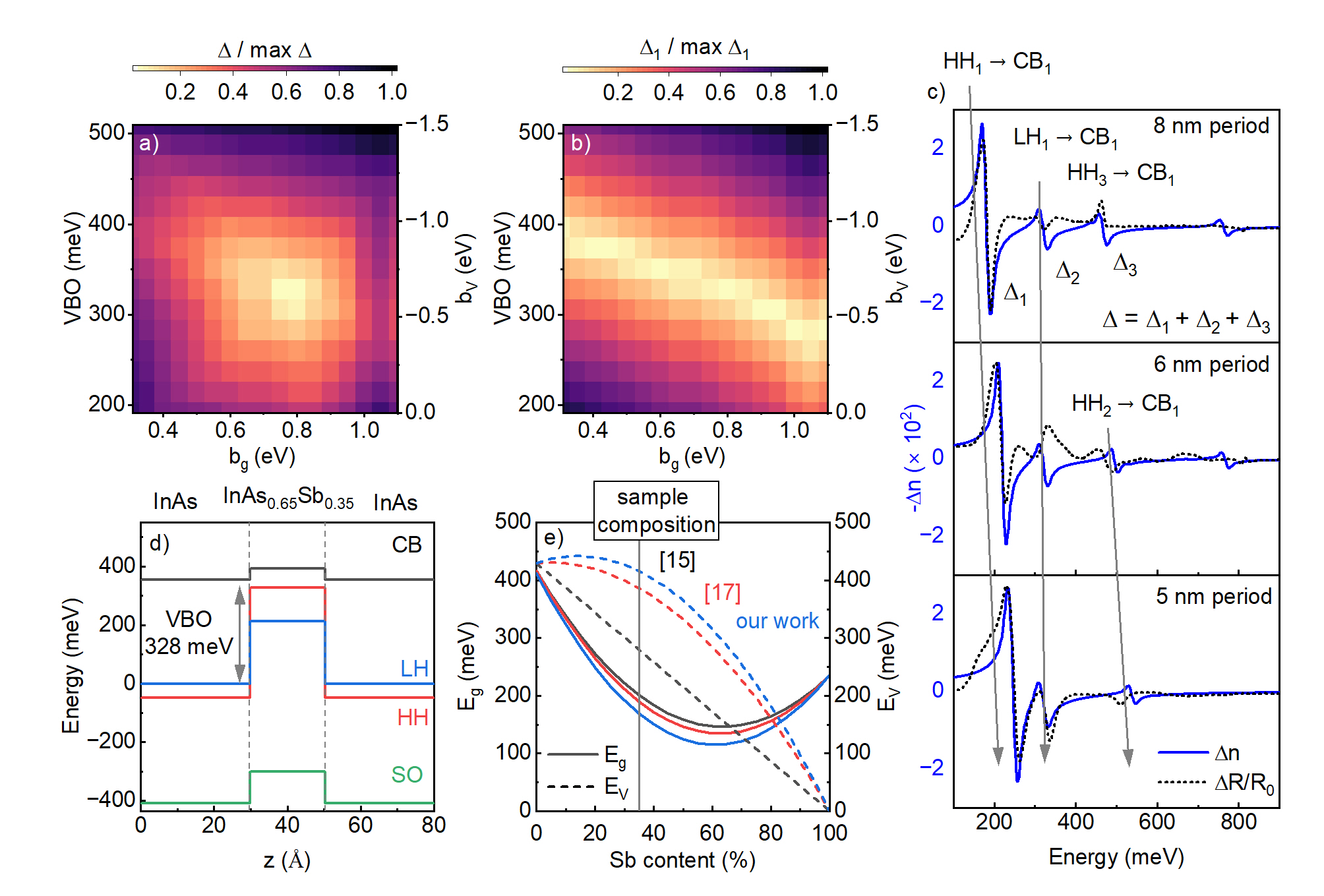}
    \caption{Comparison between the numerical calculations from 14 bulk band model and the results of photoreflectance measurements. a,b) Optimization maps for $\Delta$ (panel a) and $\Delta_{1}$ (panel b) for the 8 nm superlattice period made by plotting the difference between the resonant energies from the calculated  refractive index $\Delta$n and the energies obtained from the photoreflectance spectra $\Delta$R/R. $\Delta$n = n$(E;0)$$-$n$(E;\rho)$ was calculated for a given combination of InAsSb valence band offset b$_{\text{v}}$ and gamma gap b$_{\text{g}}$ bowing parameters, assuming a photoexcited carrier density $\rho$ equal to $2\times 10^{16}$ cm$^{-3}$. c) Comparison between $\Delta$n, calculated with the optimal parameters found in a), i.e., b$_{\text{v}} = -$0.6 eV and b$_{\text{g}} =$0.8 eV, and $\Delta$R/R for different superlattices periods $P$. Each set of parameters \{b$_{\text{g}}$, b$_{\text{v}}$\} provides a valence band offset (VBO) as shown in panel d) with the conduction (CB), heavy-hole (HH), light-hole (LH) and split-off (SO) band edges. e) Composition dependent Eq.(\ref{eq:InAsSb_energies}) for the optimal bowing parameters obtained in this work (b$_{\text{v}} = -$0.6 eV and b$_{\text{g}} =$0.8 eV), and parameters reported in the literature (Vurgaftman \textit{et al.}\cite{Vurgaftman2001} b$_{\text{v}} = 0$ eV and b$_{\text{g}} = 0.67$ eV; Manyk \textit{et al.} \cite{Manyk2018} b$_{\text{v}} =-$0.47 eV and b$_{\text{g}} = 0.72$ eV).}
    \label{fig:theory_experiment_figure_1}
\end{figure}

\begin{figure}[h]
\centering
  \includegraphics[width=0.5\linewidth]{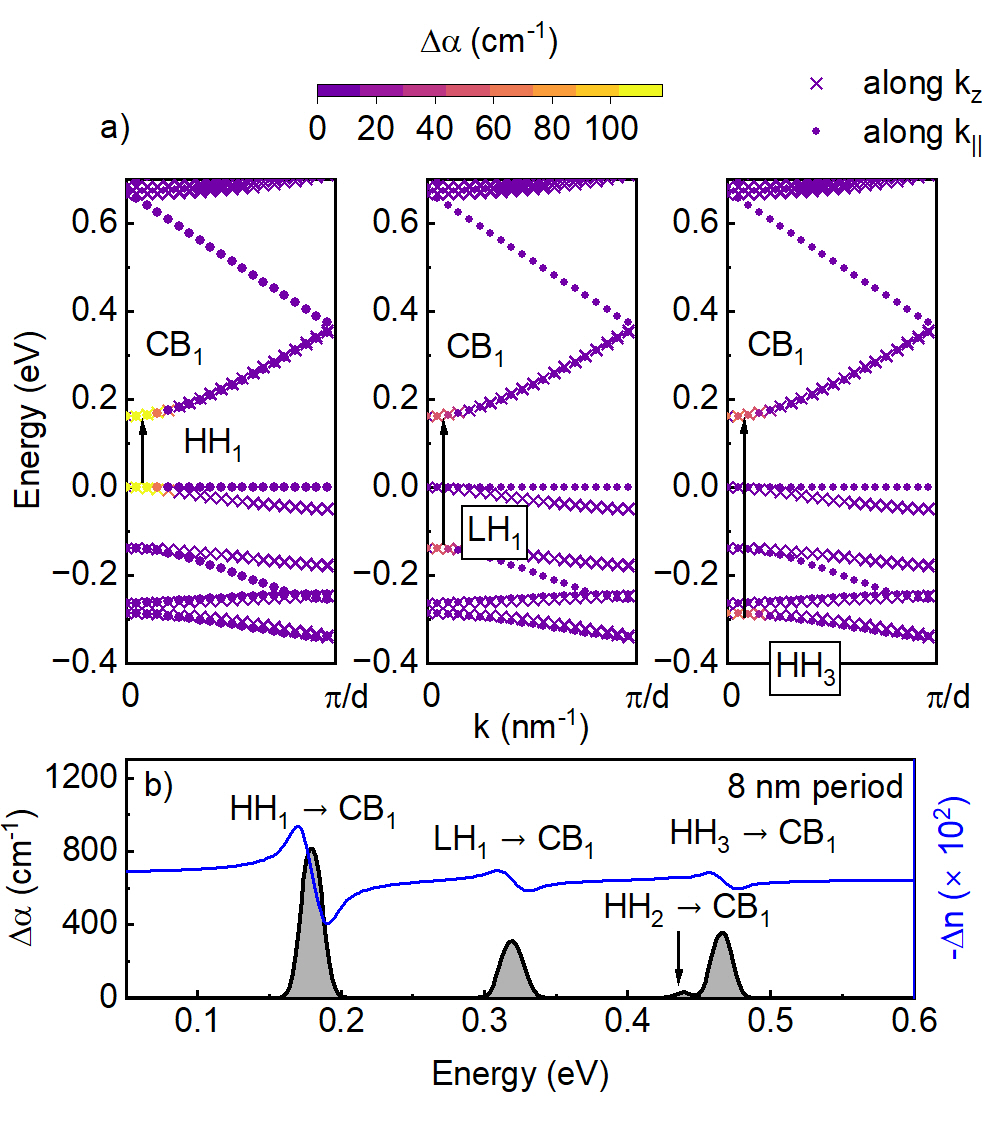}
    \caption{a) Band structure for the 8 nm period superlattice with the absorption $\Delta\alpha \equiv$ $\alpha (\text{E};0) -\alpha (\text{E};\rho)$ resolved in the k-space. \(d\) represents the superlattice period. b) Calculated refractive index \(\Delta\)n in the presence of external carriers with a density $\rho = 2\times 10^{16}$cm$^{-3}$, plotted with a sum of $\Delta\alpha$ corresponding to the states with the highest contribution to the optical resonances. k$_{z}$ and k$_{||}$ refer to directions along and perpendicular (i.e., parallel to the layer plane) to the growth direction $z$, respectively. The remaining superlattice periods were presented in \textcolor{blue}{Supplementary material}.}
    \label{fig:theory_figure_2}
\end{figure}

To gain insight into the origin of these higher energy states and at which point of superlattice band structure these transitions occur, we compared the photoreflectance spectra with the calculated refractive index curves. The curves were simulated for several different values of VBO obtained by varying the bowing parameters of InAsSb valence band offset b$_{\text{v}}$ and band gap b$_{\text{g}}$ with the results shown in Figure \ref{fig:theory_experiment_figure_1} on panels a) and b). We compared the transition energy obtained from the refractive index calculations with the photoreflectance by introducing the parameter $\Delta$ = $\Delta_{1}$ + $\Delta_{2}$ + $\Delta_{3}$, with $\Delta_{i}$ describing the difference in energies obtained between the refractive index and the photoreflectance for the $i^{th}$ (=1,2,3) transition. The transition energy from the photoreflectance is extracted using Eq. (\ref{PR modulus}) as shown in Figure \ref{fig:experimental figure 2}, while in the refractive index the energy transition is defined as the middle point between the maximum and minimum resonance energies (equivalent to the maximum value picked by the change in the absorption $\Delta\alpha$ = $\alpha(\text{E};0)$ - $\alpha(\text{E};\rho)$). All the results in Figures \ref{fig:theory_experiment_figure_1} and \ref{fig:theory_figure_2} assume layer thicknesses and Sb composition (35\%) for the superlattice as expected during an ideal growth procedure. In the \textcolor{blue}{Supplementary Material}, we presented additional analysis for the refractive index taking into account certain imperfections during growth, e.g., slightly different layer thicknesses and Sb composition deviating from the nominal values. 

The inclusion of optical transitions beyond the band gap with $\Delta$ shows a clear advantage in defining the VBO between the InAs and InAsSb layers. By comparing Figures \ref{fig:theory_experiment_figure_1}a and \ref{fig:theory_experiment_figure_1}b, it becomes clear that it is possible to minimize $\Delta$ with a set of bowing parameters resulting in a specific VBO and band gap, while for $\Delta_1$ the VBO and band gap bowing parameters seem mutually dependent with multiple sets yielding near 0 meV values.
Table \ref{tab:table1} shows a summary of the smallest values for $\Delta$ and $\Delta_{1}$, including the results for the 6 nm and 5 nm period superlattices. Among all superlattices in Table \ref{tab:table1}, the 8 nm period has the smallest value for $\Delta$ as a result of its second and third resonances in the photoreflectance having a better resolution, when compared to the results for 6 and 5 nm period samples, as shown in Figure \ref{fig:theory_experiment_figure_1}e. The lower resolution of resonances for 6 and 5 nm samples caused $\Delta$ minima to broaden to a range defined by b$_{\text{v}}$ = [$-$0.65 eV, $-$0.45 eV] and b$_{\text{g}}$ = [0.8 eV, 1 eV] which was shown in \textcolor{blue}{Supplementary Material}. Using the optimal values of b$_{\text{v}}$ and b$_{\text{g}}$ from Table \ref{tab:table1} for all the samples the calculated refractive index curves match the optical features in the photoreflectance spectra as shown in Figure \ref{fig:theory_experiment_figure_1}c.

\begin{table}[h]
\centering
\caption{Results obtained for VBO, b$_{\text{v}}$ and b$_{\text{g}}$ after optimization in Figure \ref{fig:theory_experiment_figure_1}a). The 6 and 5 nm superlattice periods have their values for $\Delta$ and $\Delta_{1}$ obtained with the same optimal parameters found for the 8 nm superlattice period. \label{tab:table1}}

\begin{tabular}{m{3em}m{3em}m{3em}m{3em}m{3em}m{3em}m{3em}}
\multirow{2}{3em}{Period}&\(\Delta\)&\(\Delta_1\)&VBO&\(\text{b}_{\text{v}}\)& E$_\text{g}$&\(\text{b}_\text{g}\)\\
&(meV)&(meV)&(meV)&(eV)&(meV)&(eV)\\
\hline
\(8~\text{nm}\)&6&1&\multirow{3}{3em}{328}&\multirow{3}{3em}{\(-0.6\)}&\multirow{3}{3em}{168}&\multirow{3}{3em}{0.8}\\
\(6~\text{nm}\)&34&2\\
\(5~\text{nm}\)&40&5\\
\end{tabular}

\end{table}

In Figure \ref{fig:theory_experiment_figure_1}e the  optimal bowing parameters, b$_{\text{v}} =-$0.6  eV and b$_{\text{g}}$ = 0.8 eV, are compared to those in refs. \cite{Vurgaftman2001, Manyk2018}. This is done by calculating the valence-band energy E$_{\text{v}}$ and band-gap energy E$_{\text{g}}$ at the gamma point as a function of the Sb composition (\%) for the InAsSb layer, using Eq.(\ref{eq:InAsSb_energies}). Note that, E$_{\text{v}}$ is referenced to the valence-band edge of InSb. For Sb = $35 \%$, our parameters result in E$_{\text{v}} = 416$ meV and E$_{\text{g}} = 168$ meV. Comparing these results with those obtained with Vurgaftaman's parameters \cite{Vurgaftman2001}, E$_{\text{v}}$ is 137 meV larger and E$_{\text{g}}$ is 32 meV smaller, respectively; Compared to T. Manyk's \cite{Manyk2018}, which are based on an experimental study of superlattices with various Sb content \cite{Svensson2012}, E$_{\text{v}}$ is 30 meV larger and E$_{\text{g}}$ is 21 meV smaller. 

 In order to identify the originating states resulting in photoreflectance resonances in Figure \ref{fig:theory_experiment_figure_1}c, we plotted the change in the absorption $\Delta\alpha \equiv$ $\alpha (\text{E};0) -\alpha (\text{E};\rho)$ with the band structure, as shown in Figure \ref{fig:theory_figure_2}. Three panels on Figure \ref{fig:theory_figure_2}a clearly show that the transitions assigned by, respectively, HH$_{1} \rightarrow$CB$_{1}$, LH$_{1} \rightarrow$CB$_{1}$ and HH$_{3} \rightarrow$CB$_{1}$, occur predominantly close to k$_{z} = 0$. For 8 nm period the third resonance for $\Delta$n in Figure \ref{fig:theory_figure_2}b has a small additional contribution from the transition HH$_{2} \rightarrow$CB$_{1}$. This transition has the highest contribution to the third resonance for 6 and 5 nm periods. The corresponding figures for the remaining samples were shown in the \textcolor{blue}{Supplementary Material}.

\section{CONCLUSION}
In this work, three InAs/InAs$_{0.65}$Sb$_{0.35}$ superlattices with different periods were investigated using photoluminescence and photoreflectance measurements. The photoluminescence from 80 K to 10 K shows an unexpected progression for the emission energy for all the samples that is not in agreement with the Varshini-predicted bandgap. We attribute this behavior to localized states, such as traps caused by indium vacancies. This hypothesis is supported by the photoreflectance data, showing that its main resonance happens at the energy expected to be the bandgap for the superlattice. 

The photoreflectance is further compared with the change in refractive index, which is calculated using a 14×14 \textbf{k}$\cdot$\textbf{p} model for various InAsSb bowing parameters of the valence band offset, b$_{\text{v}}$, and the bandgap b$_{\text{g}}$. We identify the bands in the superlattices associated with the transitions in the photoreflectance, and show that higher-energy optical transitions (beyond the fundamental band-gap transition) strongly impact the choice of the valence band offset between InAs and InAsSb. Our comparison yields a InAs$_{0.65}$Sb$_{0.35}$ valence band offset (in relation to InAs) of 328 meV, obtained using the bowing parameters b$_{\text{v}} =-$0.6 eV and b$_{\text{g}}=$0.8 eV.

\section*{ACKNOWLEDGEMENTS}
This work has been funded by the German Federal Ministry of Research, Technology, and Space (BMFTR) within the joint research project RIBKD  (13N16964).  J.~Z.  was financially supported by Marie Sklodowska-Curie  Grant No. 956548.
\section*{SUPPLEMENTARY MATERIALS}
See Supplementary Material for more information on the mathematical formulas used for calculations of the refractive index, extended optimization study taking into account the imperfections of the growth procedure and the results of the optimization for the superlattices with 6 and 5 nm period.
\section*{DATA SHARING POLICY}
The data that support the findings of this study are available from the corresponding author upon reasonable request.
%Bibliography
\bibliographystyle{unsrt}  
\bibliography{references}

\end{document}